\documentclass[prb,aps,twocolumn,showpacs,preprintnumbers,amsmath,amssymb,superscriptaddress]{revtex4-1}
\usepackage[usenames,dvipsnames]{xcolor}

\usepackage{float}
\usepackage{amsfonts}
\usepackage[english]{babel}
\usepackage[T1]{fontenc}
\usepackage{times}
\usepackage{mathrsfs}
\usepackage{graphicx}
\usepackage{dcolumn}
\usepackage{bm}
\usepackage[colorlinks,bookmarks=false,citecolor=blue,linkcolor=red,urlcolor=blue]{hyperref}
\usepackage{natbib}

\newcommand{\Eref}[1]{Eq.~(\ref{#1})}

\newcommand{\Fref}[1]{Fig.~\ref{#1}}
\newcommand{\Sref}[1]{Sec.~\ref{#1}}
\newcommand{\tr}{\mathrm{Tr}}
\newcommand{\etal}{{\it et al.}~}
\newcommand{\nmax}{n_\mathrm{max}}

\begin{document}

\title{Entanglement Spectroscopy using Quantum Monte Carlo}

\author{Chia-Min Chung}
\email{chiaminchung@gmail.com}
\affiliation{Institute for Theoretical Physics, University of Innsbruck, A-6020 Innsbruck, Austria}
\affiliation{Department of Physics, National Tsing Hua University, Hsinchu 30013, Taiwan}

\author{Lars Bonnes}
\email{lars.bonnes@uibk.ac.at}
\affiliation{Institute for Theoretical Physics, University of Innsbruck, A-6020 Innsbruck, Austria}

\author{Pochung Chen}
\affiliation{Department of Physics, National Tsing Hua University, Hsinchu 30013, Taiwan}
\affiliation{Frontier Research Center on Fundamental and Applied Sciences of Matters, 
National Tsing Hua University, Hsinchu 30013, Taiwan}

\author{Andreas M. L\"{a}uchli}
\affiliation{Institute for Theoretical Physics, University of Innsbruck, A-6020 Innsbruck, Austria}

\date{\today}

\begin{abstract}
We present a numerical scheme to reconstruct a subset of the entanglement spectrum of quantum many body systems 
using quantum Monte Carlo. The approach builds on the replica trick to evaluate particle number resolved
traces of the first $n$ of powers of a reduced density matrix. From this information we reconstruct $n$ entanglement spectrum levels
using a polynomial root solver. We illustrate the power and limitations of the
method by an application to the extended Bose-Hubbard model in one dimension where we are able to resolve the
quasi-degeneracy of the entanglement spectrum in the Haldane-Insulator phase. In general the method is able to
reconstruct the largest few eigenvalues in each symmetry sector and typically performs  better when the eigenvalues are not too
different.
\end{abstract}

\maketitle

\section{Introduction}

The field of quantum many body systems has been boosted tremendously by the investigation of quantum 
information motivated quantities such as von Neumann or Renyi entanglement entropies applied to
strongly correlated quantum matter~\cite{amico08}. Important achievements are the discovery 
of area laws controlling the entanglement properties of ground states of local gapped
Hamiltonians~\cite{Eisert2010}, that entanglement entropies encode topological properties of 
matter~\cite{kitaev06,levin06,Isakov2011a,Jiang2012}, or that the logarithmic corrections of the area law 
in one-dimensional critical systems are governed by the central charge of the 
underlying conformal field theory (CFT)~\cite{Holzhey1994,Vidal2003,Calabrese2004}. 

More recently the entanglement spectrum - i.e. the negative logarithm of the eigenvalues of a reduced
density matrix - has been proposed as a novel tool to obtain insightful information beyond the information 
content of individual entanglement entropies~\cite{Li2008}. More specifically let us consider a bipartition of a 
system into two complementary parts $A$ and $B$ and the corresponding Schmidt decomposition of a 
wave function $|\psi\rangle$:
\begin{equation}
  \label{Schmidt}
  |\psi\rangle = \sum_i e^{-\xi_i/2} |\psi_i^A\rangle \otimes |\psi_i^B\rangle,
\end{equation}
where $|\psi_i^A\rangle$ and $|\psi_i^B\rangle$ are orthonormal vectors in subsystems $A$ and $B$ respectively, and
$\xi_i$ are the so called entanglement spectrum levels, related to the eigenvalues $\lambda_i$ of the corresponding reduced density 
matrix through the relation $\xi_i= - \ln \lambda_i$. It was shown in Ref.~\onlinecite{Li2008} that for fractional quantum Hall states 
the entanglement spectrum $\xi_i$ arranges in close analogy to the energy spectrum of a physical edge. This observation sparked 
a lot of activity exploring the physical information contained in the entanglement spectrum e.g. in topological 
quantum matter~\cite{Li2008,
brayali2009,Regnault2009,
Lauchli2010,Thomale2010a,Thomale2010b,Pollmann2010,turner2010,kargarian2010,prodan2010,fidkowski2010,
papic2011,dubail11,hughes2011,fidkowski2011,
Qi2012,turner2012}
, in continuous symmetry breaking states~\cite{Metlitski2011,Alba2012b} or 
in one-dimensional critical systems described by a CFT~\cite{Calabrese2008,Pollmann2009,DeChiara2012,Lauchli2013}.

So far numerical entanglement spectra have been obtained  
mainly using wave function based numerical approaches such as Exact Diagonalization~\cite{Li2008,Regnault2009,Lauchli2010,Thomale2010a,Thomale2010b}, 
Density Matrix renormalization Group (DMRG)~\cite{Pollmann2010,DeChiara2012,Lauchli2013,
okunishi99,Deng2011,Deng2013,jiang2010} or Tensor Network states~\cite{Cirac2011}. 
Motivated by recent advances in measuring Renyi entropies in Quantum Monte Carlo methods (QMC)~\cite{melko10,hastings10,humeniuk12} we want to explore in this paper to what extent current 
Quantum Monte Carlo methods are suitable to reconstruct entanglement spectra.

The basic idea is to measure the trace of the first $n$ powers of a reduced density matrix using a replica approach, and then to infer $n$ eigenvalues of the reduced density matrices based on the measured moments. 
This problem is a particular instance of the (truncated) Hausdorff problem~\cite{hausdorff21,shohat70} and known to be an ill-conditioned problem, similar to the analytic continuation problem arising in the Monte Carlo evaluation of spectral functions. An additional complication is the presence of statistical uncertainties of the Renyi entropies similar to the situation of a statistical Hausdorff problem considered in Ref.~\onlinecite{ngoc08}.

With these limitations in mind it is our goal to explore how far one can go with current technologies. While the full reconstruction of the entanglement spectrum is clearly out of reach, a particle number resolved measurement technique allows us to track roughly the lowest few entanglement spectra levels per sector with reasonable effort, provided the eigenvalues are of similar magnitude.

We illustrate the approach by simulating the one-dimensional extended Bose-Hubbard model focusing on the Mott insulator and the Haldane insulator.
We demonstrate that our QMC approach is able to detect the quasi-degeneracy of the entanglement spectrum in the Haldane insulator versus the non-degeneracy in the Mott insulator.

\section{method}
\subsection{QMC Implementation.}
\begin{figure}
\includegraphics[width=0.9\columnwidth]{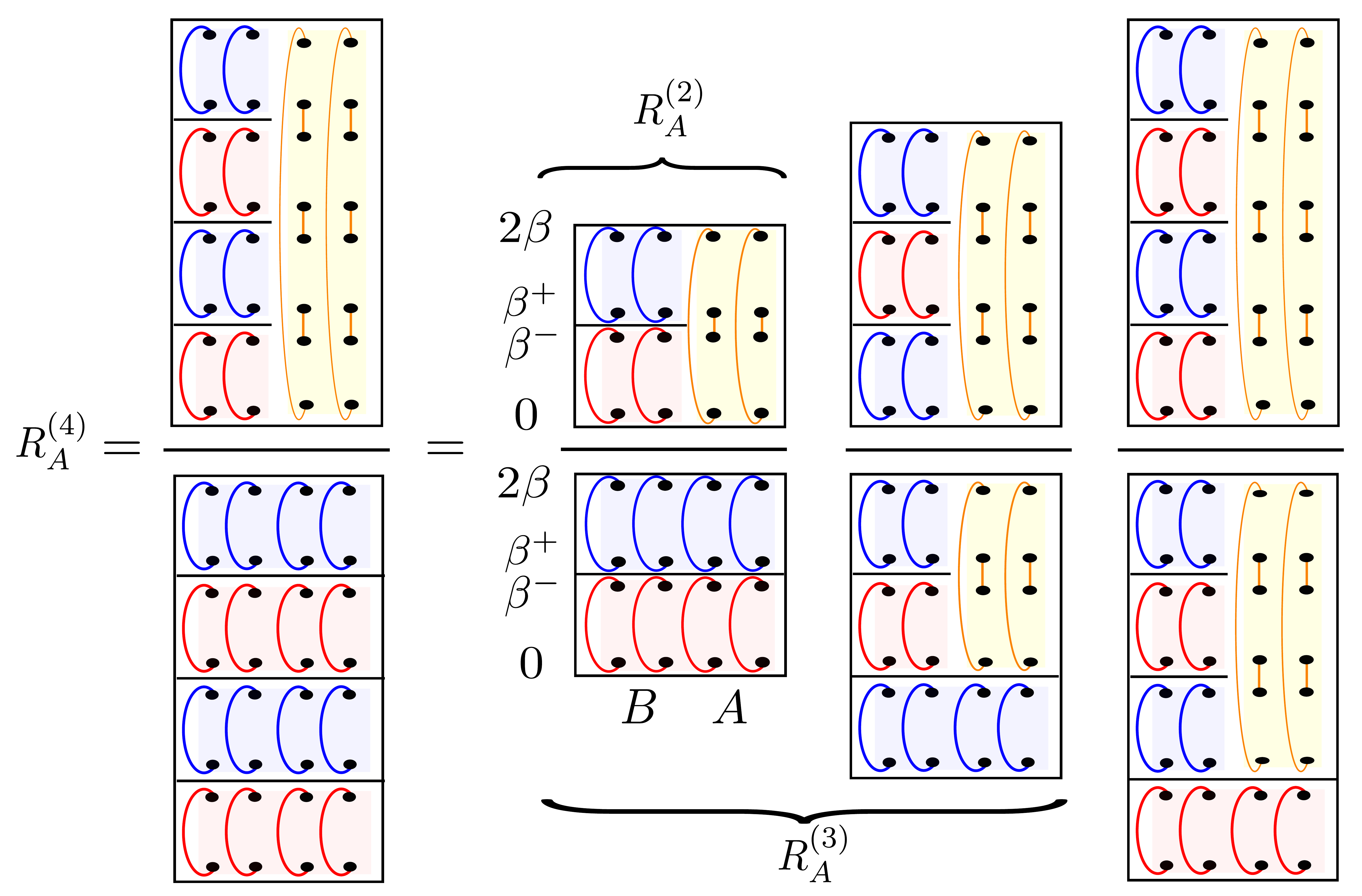}
\caption{(Color online) 
Illustration of the world line connectivity required for the calculation of $R_A^{(4)}$.
The right hand side illustrates the {\it imaginary time increment trick} used to estimate $R_A^{(n)}$.
}
\label{fig:wld}
\end{figure}

The $n$th Renyi entropy, $S^{(n)}_A$, can be accessed in (quantum) Monte Carlo using the replica trick~\cite{melko10} that links the $n$th moment of the reduced density matrix to a ratio of two partition functions,
\begin{equation}
  R_A^{(n)} \equiv \tr \rho_A^n = \frac{Z_A^{(n)}}{Z^n},
 \label{eq:R}
\end{equation}
and $S_A^{(n)}= 1/(1-n) \log R_A^{(n)}$.
$Z=\tr \exp(-\beta H)$ denotes the usual partition function for a system at temperature $T=1/\beta$ and $Z_A^{(n)}$ is the partition function living on a $n$-sheeted Riemann surface of temporal extent $n \beta$ with $n-1$ equal-time branch-cuts at times $m \beta$ ($m=1,..,n-1$) along the extent of subsystem $A$.
Here we use a continuous time worm algorithm~\cite{prokofev98a,prokofev98b,pollet07b} that operates in the path integral representation of the partition functions and allows for an efficient sampling of $R_A^{(n)}$ when the normal world-line update is supplemented with a global update scheme devised in Ref.~\onlinecite{humeniuk12}.
Starting from $Z^n$, the global update tries to connect the world lines along the branch cuts within $A$.
If the world lines at $m\beta^-$ can be connected to $m\beta^+$ (note the periodicity in imaginary time), the world line sheets are transfromed to this non-trivial geometry $Z_A^{(n)}$, as illustrated in \Fref{fig:wld}.
An inverse move tries to recut the world lines at all $m\beta^-$ and $m\beta^+$.
$R_A^{(n)}$ is estimated by simply counting how often the system is in one or the other world line configuration.

As $A$ grows the transition probability in the global update decreases rapidly and the simulation becomes rather inefficient. 
By expanding \Eref{eq:R} as $[Z_{A_1}^{(n)}/Z^{(n)}] [Z_{A_2}^{(n)}/Z_{A_1}^{(n)}] ... [Z_{A}^{(n)}/Z_{A_k}^{(n)}]$ with $A_1 \subset A_2 \subset \ldots\subset  A_k \subset A$ , large subsystem sizes can be accessed by successively growing the subsystem and evaluating each partition function ratio in square brackets separately~\cite{hastings10}.

Accessing higher Renyi entropies poses a similar challenge since the number of branch cuts increases.
To cure this inefficiency, we conceive an \textit{imaginary time increment trick} by realizing that the partition function ratio can be rewritten as
\begin{equation}
  R_A^{(n)} = \frac{Z_{A}^{(n-1)}}{Z^{n-1\phantom{)}}_{\phantom{A}}} \frac{Z_{A}^{(n)}}{ZZ_A^{(n-1)}} = R_A^{(n-1)}\frac{Z_A^{(n)}}{ZZ_A^{(n-1)}}.
\label{eq:ZZ}
\end{equation}
Thus, the $n$th Renyi entropy  requires the calculation of $R_A^{(n-1)}$ as well as the evaluation of a partition function ratio between two geometries with a \textit{single} branch cut, as illustrated in \Fref{fig:wld}.
This improved scheme allows us to access Renyi entropies as high as $n=4$ and can readily be generalized to different implementations. 
Although it is possible to access also higher Renyi entropies efficiently with this update scheme we restricted ourselves to the calculation of up to the fourth moment since including higher moments will not improve the reconstruction of the entanglement spectrum given the current accuracy of our data as it will be detailed in the following section.

\subsection{Reconstructing the Entanglement Spectrum.}
\label{sec:MethExtr}
\begin{figure}
\includegraphics[width=0.9\columnwidth]{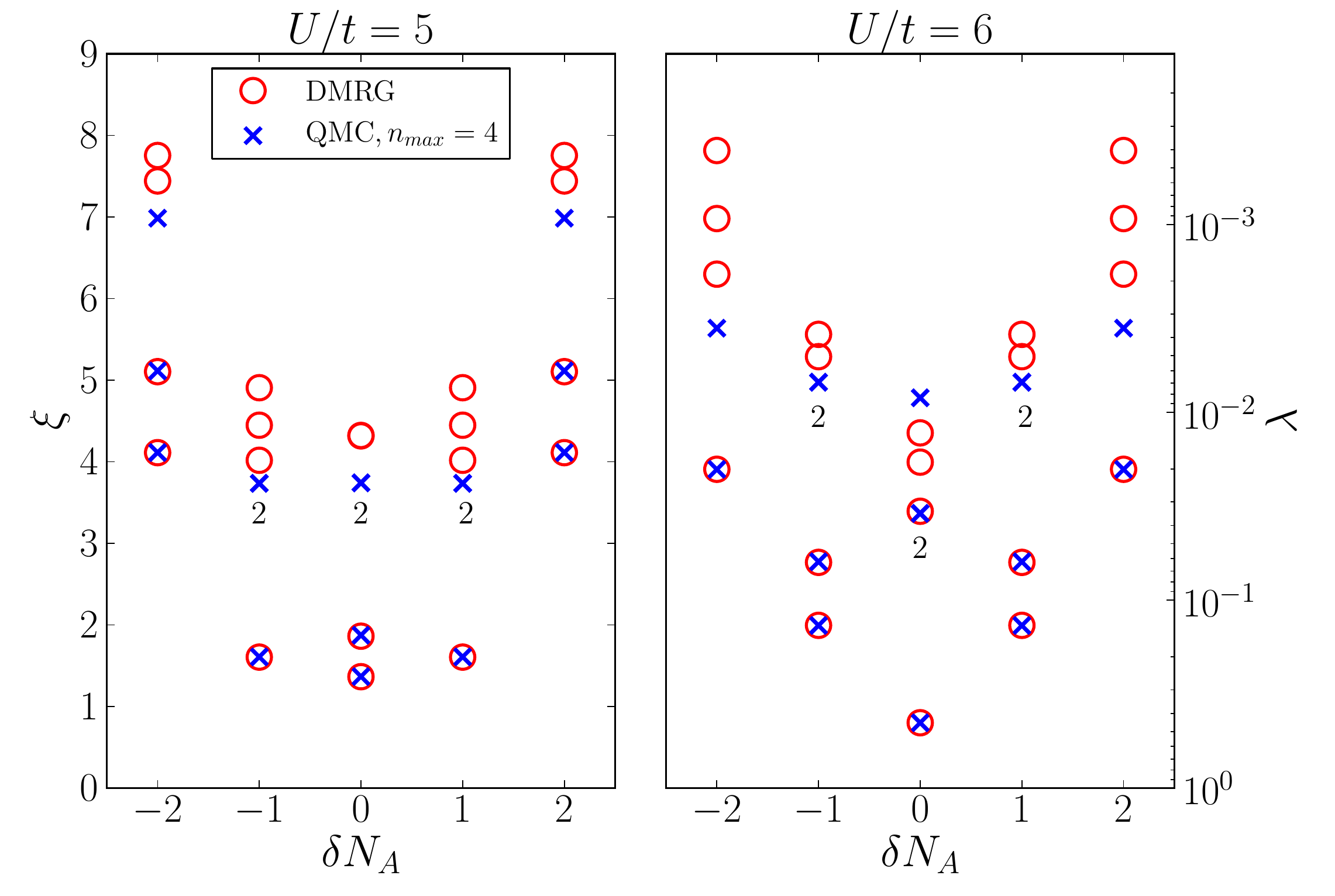}
\caption{(Color online) 
Lower part of the particle number resolved entanglement spectrum of a block of length $L_A=24$ in a periodic chain with $L=48$ of an extended Bose-Hubbard model at unit filling with $V/t=3.3$ and $U/t=5$ (HI, left panel) and $U/t=6$ (MI, right panel) obtained using $\nmax=4$.
An (artificial) degeneracy is indicated by numbers below the data points. Only the lowest four DMRG levels per sector are shown.
}
\label{fig:xi}
\end{figure}

\begin{figure*}
\includegraphics[width=0.3\linewidth]{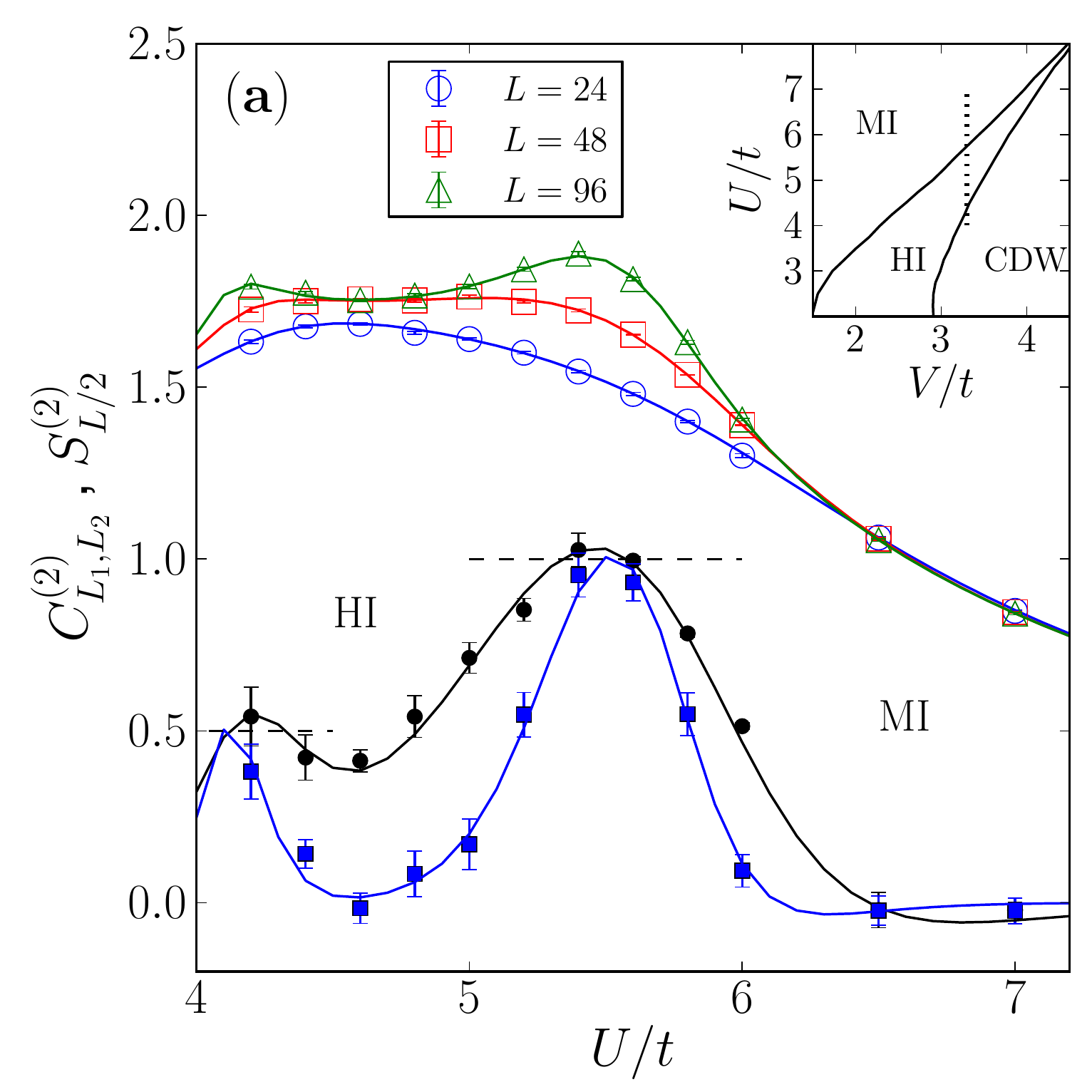}
\includegraphics[width=0.62\linewidth]{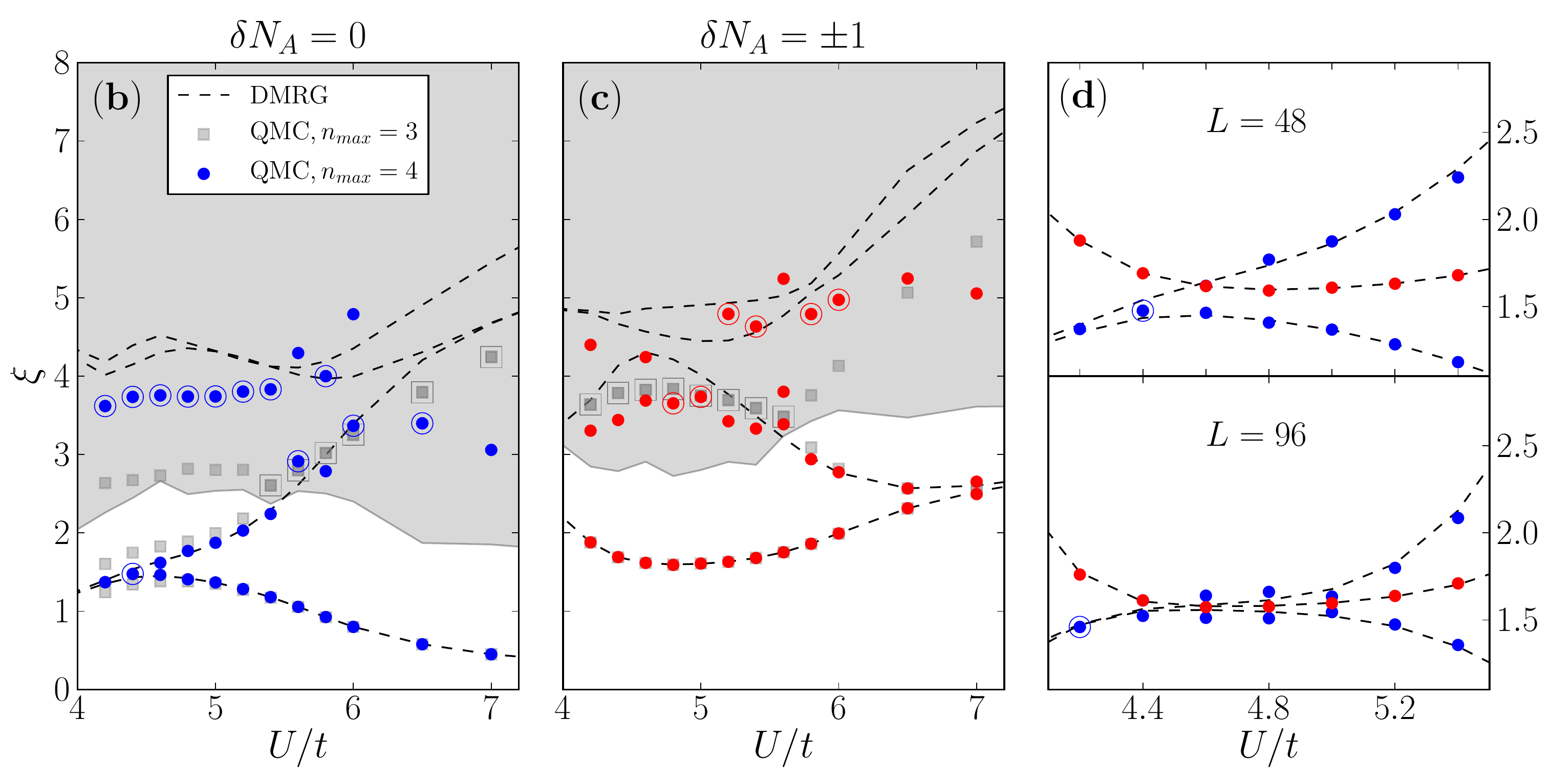}
\caption{(Color online)
(a) Renyi entropies $S_{L/2}^{(2)}$ (open symbols) and the central charge estimator $C^{(2)}_{L_1,L_2}$ for $(L_1,L_2)=(24,48)$ (filled black dots) and $(48,96)$ (filled blue square) as a function of $U/t$ for $V/t=3.3$. The continuous lines denote reference DMRG results. 
The inset shows part of phase diagram from Ref.~\onlinecite{Deng2011}, where the dotted line indicates the parameter region we focus on.
(b),(c) The lower part of ES on $\delta N_A=0$ and $\pm 1$ as a function of $U$, gotten by $\nmax=3$ and $\nmax=4$. 
The open symbols indicate the artificial two-fold degenerate points. The gray shaded area is the regime $\xi>-1/4\log\delta R_A^{(4)}$ where the ES cannot be resolved accurately.
(d) The panel displays lowest two entanglement levels for $\delta N_A=0$ (blue dots) and $\delta N_A=\pm 1$ (red dots) in the HI for $L=48$ (from panels (b) and (c)) and $L=96$ showing the closing of the gap in the entanglement spectrum with increasing system size.
The dashed lines in panels (b) to (d) denote the reference DMRG data.
}
\label{fig:Sxi}
\end{figure*}

The knowledge of all integer Renyi entropies in principle provides all the information required to extract the entanglement spectrum~\cite{Calabrese2008} by, for instance, solving a large system of coupled non-linear equations.
We explored several entanglement spectrum reconstruction approaches including an optimization techniques over the $n$ eigenvalues. Since this operation has to be performed many times in the binning analysis
of the Monte Carlo time series, 
we chose to rely on the recent work by Song \etal~\cite{song12}, which presented a nice way to relate the $R_A^{(n)}$ quantities to the characteristic polynomial of $\rho_A$.~\footnote{The data obtained from the optimization procedure and the root solver is consistent.} In our setup we can only
get accurate data up to certain $n$, so we simply approximate the full characteristic polynomial by an order $n$ truncation. Finding the roots of a polynomial with noisy coefficients can, in general, be challenging 
(see e.g. Ref.~\onlinecite{guillaume89}), for example the roots can sometimes shift away from the real axis and cluster in pairs, thereby acquiring a small imaginary part that will be omitted.~\footnote{Not only the statistical uncertainty of the coefficients but also the mere truncation procedure can lead to imaginary parts in the root finding process.}
This does not, however, affect the normalization condition $\sum_i \lambda_i=1$ since the imaginary parts cancel.

The main difficulty in reconstructing the spectrum of the reduced density matrix is that the eigenvalues $\lambda$ can extend over several orders of magnitude.
Higher moments, on the other hand, are dominated by the largest eigenvalue $\lambda_1$, i.e. $R_A^{(n)}=\lambda_1^n (1 + [\lambda_2/\lambda_1]^n + ...)$, 
providing a threshold for the resolvable eigenvalues given an error on $R_A^{(n)}$.

An  improvement of the reconstruction procedure is to take advantage of the block structure of the density matrix that stems from particle number conservation (or any other quantum numbers 
that are easily accessible in QMC). To be more specific, we calculate the moments of $\rho_A$ in each particle number sector labelled by $\delta N_A = N_A - \bar{N}_A$ ($N_A$ is the number of particles in block $A$ and $\bar{N}_A$ is the average particle number) separately by binning the measurements into the respective sectors. 
We combine the data for symmetric blocks with $\delta N_A$ and $-\delta N_A$ to improve the accuracy of our data.~\footnote{This is an exact symmetry for $L_A=L/2$ and we checked explicitly that the data for blocks with $\pm \delta N_A$ agrees within the error bars.}

The power of our approach is illustrated in \Fref{fig:xi}, where we show entanglement spectra for two different parameter sets in a one-dimensional extended Bose Hubbard model.
The model as well as the physics of the spectra will be discussed in Sec.~\ref{sec:bosehubbard} below. Here we just want to highlight the good agreement between the QMC reconstruction of the
lowest entanglement spectrum level in each sector (corresponding to the largest eigenvalues $\lambda$) and the reference DMRG results. Furthermore also the second entanglement
spectrum level in each sector is rather accurate, provided the $\xi$ values are not separated by a large gap from the first level in the same sector. 

\section{Numerical results}
\label{sec:bosehubbard}
For a more thorough application we now turn to the extended Bose-Hubbard model here given in terms of standard bosonic operators as
\begin{eqnarray}
  \label{eq:H}
  H &=& -t\sum_i \left( b_i^\dagger b_{i+1} + \mathrm{h.c.} \right) 
+ \frac{U}{2} \sum_i n_i \left( n_i-1\right) \nonumber\\
   &+& V\sum_i n_i n_{i+1}.
\end{eqnarray}
Recent analytical~\cite{torre06,berg08} and numerical~\cite{amico10,Deng2011,rossini12,batrouni13} studies revealed that its phase diagram features a topologically non-trivial Haldane insulating (HI) phase besides more
conventional phases such as a Mott insulator (MI) and a charge density wave phase (CDW).
The HI is closely linked to the Haldane phase in spin $S=1$ chains that does not break a local symmetry but has non-local string order and a characteristic 
degeneracy of the entanglement spectrum protected by a set of symmetries~\cite{Pollmann2010,Pollmann2012}.

We restrict ourselves to the case of unit filling and periodic boundary conditions (PBC) and sketch the phase diagram, obtained from Ref.~\onlinecite{rossini12}, in the inset of \Fref{fig:Sxi}(a).
The simulations are performed along the line $V/t=3.3$ where the system can be tuned from the CDW at lower $U$ to the HI at $U/t \approx 4.15$.
Upon further increasing the on-site interaction, one eventually enters the MI at $U/t \approx 5.55$.

\subsection{Locating the Phase Transitions}

As a preparatory step we investigate the behavior of the second Renyi entropy of a half system block $L_A=L/2$ across the two phase transitions. In a gapped phase
the entropy of a block saturates as a function of the block size by virtue of the strict area law. At a quantum critical point with conformal symmetry however, the entanglement entropies exhibit a
logarithmic correction, whose coefficient is directly related to the central charge $c$:
\begin{equation}
S^{(n)}_{l_A}(U_c) = \frac{c}{6}(1+\frac{1}{n}) \log l'_A + c'_n\ ,
\label{eq:logDiv}
\end{equation}
where $l'_A$ is the chord length of block $A$ and $c'_n$ is a non-universal constant. When considering the entropy increment
upon increasing the system and block size~\cite{laeuchli08}:
\begin{equation}
  \label{eq:dS}
  C_{L_1,L_2}^{(2)} = \frac{4}{\log L_2/L_1}\left[S^{(2)}_{L_2/2}-S^{(2)}_{L_1/2}\right],
\end{equation}
this quantity displays a peak at the location of the quantum phase transition and its peak value corresponds to the central charge of the transition. In the 
gapped phase $C_{L_1,L_2}^{(2)}$ tends to zero with increasing system sizes. In \Fref{fig:Sxi}(a) we find two distinct peaks with $c=1/2$ (CDW-HI) and $c=1$ (HI-MI) 
in accordance with the expected presence of an Ising and a Gaussian critical point respectively~\cite{chen03}. Furthermore our QMC data -- also for higher 
Renyi entropies (not shown) -- agrees perfectly with the reference DMRG results.

\subsection{Entanglement Reconstruction}
We now turn towards the analysis of the entanglement spectrum as a function of $U/t$ for fixed $V/t=3.3$, focussing on the MI and HI phases.
In insightful papers Pollmann \etal~\cite{Pollmann2010,Pollmann2012} realized that the HI phase can be characterized by a particular degeneracy
structure of the entire entanglement spectrum. In the PBC setup we consider, there are two distinct boundaries separating the two parts of the system
and the degeneracy of the lowest lying multiplet is four in this case, with a quantum number structure as shown in the left panel of~\Fref{fig:xi}, i.e.~two levels at
$\delta N_A=0$ and one level each at $\delta N_A=\pm1$.

In \Fref{fig:Sxi}(b) and (c) we now display the $U/t$ dependence of the QMC reconstructed ES based on the method outlined in \Sref{sec:MethExtr}. 
The gray squares denote data reconstructed using $\nmax=3$, whereas the colored dots correspond to $\nmax=4$.
In panel \Fref{fig:Sxi}(d) we display the lowest two levels for $\delta N_A=0$ and $\delta N_A=\pm 1$ for $L=48$ and $96$ in the HI. 
One can clearly see that the lowest four levels become closer for increasing system size, eventually leading to the four-fold degeneracy in the thermodynamic limit.
In fact, the QMC calculations nicely capture this feature and agree with the DMRG results.

Taking either three or four moments of the density matrix, the lowest two levels of the entanglement spectrum can be captured quite well. 
Although the $\nmax=4$ data for the third level in the $\delta N_A=0$ sector agrees with the DMRG results qualitatively, the quantitative agreement is not as good as for the lower levels.
This effect is readily understood as a small and finite number of entanglement levels will necessarily lead to a deformation of the spectrum -- some of the $\xi$s are typically shifted downwards because the complete trace has to be incorporated in very few levels. 

A second observation is that the entanglement gap in the respective particle number sectors has a substantial influence on the quality of the obtained spectrum.
The separation of the $\xi$ values for large $U$ increases the domination of the first level and hence requires a much better resolution as compared to the Haldane phase where the lowest two levels are almost degenerate.

We find that we can actually resolve entanglement levels only below a threshold $\xi \lesssim -1/n \log[\delta R_A^{(n_\mathrm{max})}]$ given by the error on the highest moment.
This threshold, that is also shown in \Fref{fig:Sxi}, stems from the observation that the overall precision of the moments has to be comparable or better than the summands one tries to resolve and also incorporates the fact that smaller eigenvalues of the density matrix are suppressed exponentially, i.e. like $(\lambda_j/\lambda_1)^n$, for increasing $n$.~\footnote{This does {\it not} imply that reducing $n_\mathrm{max}$ will yield a better resolution of the entaglement spectrum since the absolute precision due to the truncation of the spectrum will become worse.}
Above the error threshold the spectrum easily acquires false degeneracies, as discussed previously, or some of the reconstructed eigenvalues can become purely imaginary and are thus omitted.

\section{Summary and Discussion}
In this paper we have presented an approach to reconstruct the entanglement spectrum of quantum many body systems by measuring particle number resolved higher moments of reduced density matrices in QMC.
To effectively calculate larger $n$ Renyi entropies, an enhanced increment scheme is presented that greatly improves the global topology updates of the world lines.
We are in fact able to recover the low lying degenerate levels in the HI insulator and show that this degeneracy does not persist in the MI phase where, however, our method fails to quantitatively capture the higher-lying part of the spectrum of the MI because of the large separation of levels.
 
Our approach is not only limited to the QMC methods described in this paper. Similar procedures can potentially
be applied to Variational Monte Carlo (VMC) evaluation of trial wavefunctions, going beyond the 2nd Renyi entropy calculations
performed so far~\cite{zhang2011a,zhang2011b,zhang2012a,zhang2012b}, and thus potentially access a few of the largest eigenvalues of the reduced density matrices. 

Another interesting aspect is that using the protocol put forward in Ref.~\onlinecite{daley12} it seems possible to both obtain higher $R^{(n)}$ as well as particle number resolved traces $R^{(n)}[\delta N_A]$, so that a partial reconstruction of the entanglement
spectrum in experiments on large bosonic systems might become possible in the future.

\acknowledgments
We acknowledge support by the Austrian Science Fund (FWF) through the SFB FoQuS (FWF Project No.~F4018-N23).
We acknowledge financial support and allocation of CPU time from NSC and NCTS Taiwan. This work was supported by 
the Austrian Ministry of Science BMWF as part of the UniInfrastrukturprogramm of the Forschungsplattform Scientific Computing 
at LFU Innsbruck.


%

\end{document}